\documentclass{ws-ijmpa}
\usepackage{cite}
\usepackage{graphicx}
\usepackage{color}

\newcommand{\Dp}{D^{\prime}}
\newcommand{\Fp}{F^{\prime}}

\newcommand{\ag}{a_{3g}}
\newcommand{\aga}{a_{\gamma}}
\newcommand{\ac}{a_c}

\newcommand{\psp}{\psi^{\prime}}

\newcommand{\jpsi}{J/\psi}

\newcommand{\EE}{e^+e^-}

\newcommand{\kskl}{K^0_SK^0_L}

\newcommand{\nnb}{n\overline{n}}

\newcommand{\BBb}{B\overline{B}}

\newcommand{\ppb}{p\overline{p}}

\newcommand{\LLb}{\Lambda \overline{\Lambda}}

\newcommand{\SSbz}{\Sigma^0 \overline{\Sigma}^0}
\newcommand{\SSbp}{\Sigma^+ \overline{\Sigma}^-}
\newcommand{\SSbm}{\Sigma^- \overline{\Sigma}^+}
\newcommand{\XXbz}{\Xi^0 \overline{\Xi}^0}
\newcommand{\XXbm}{\Xi^- \overline{\Xi}^+}
\newcommand{\SzLb}{\Sigma^0 \overline{\Lambda}}

\newcommand{\SbzL}{\overline{\Sigma}^0 \Lambda}

\newcommand{\VP}{1^-0^-}
\newcommand{\PP}{0^-0^-}

\newcommand{\beq}{\begin{equation}}
\newcommand{\eeq}{\end{equation}}
\newcommand{\beqn}{\begin{eqnarray}}
\newcommand{\eeqn}{\end{eqnarray}}
\newcommand{\beqns}{\begin{eqnarray*}}
\newcommand{\eeqns}{\end{eqnarray*}}
\newcommand{\bfg}{\begin{figure}}
\newcommand{\efg}{\end{figure}}
\newcommand{\bitm}{\begin{itemize}}
\newcommand{\eitm}{\end{itemize}}
\newcommand{\bnum}{\begin{enumerate}}
\newcommand{\enum}{\end{enumerate}}
\newcommand{\btbl}{\begin{table}}
\newcommand{\etbl}{\end{table}}
\newcommand{\btbu}{\begin{tabular}}
\newcommand{\etbu}{\end{tabular}}
\def\eref#1{(\ref{#1})}
\def\Journal#1#2#3#4{{#1} {\bf #2}, #3 (#4)}

\def\PLB{Phys. Lett. B}

\def\PRL{Phys. Rev. Lett.}
\def\PRD{Phys. Rev. D}

\def\HEPNP{HEP \& NP}

\def\prd#1#2#3 {{~Phys. Rev. D {#1}, #2 (#3) }}  
\def\plb#1#2#3 {{~Phys. Lett. B {#1}, #2 (#3) }}  

\newsavebox{\arrect}
\savebox{\arrect}(0,0){%
\setlength{\unitlength}{0.5cm}
\thicklines
\put(-4,1){\line(1,0){8.0}}
\put(4,1){\line(0,-1){2.0}}
\put(4,-1){\line(-1,0){8.0}}
\put(-4,-1){\line(0,1){2.0}}
\put(0,-1){\vector(0,-1){1.0}}
}

\newsavebox{\arrhomb}
\savebox{\arrhomb}(0,0){%
\setlength{\unitlength}{1cm}
\thicklines
\put(0,0.5){\line(-2,-1){1.0}}
\put(0,0.5){\line(2,-1){1.0}}
\put(0,-0.5){\line(-2,1){1.0}}
\put(0,-0.5){\line(2,1){1.0}}
\put(0,-0.5){\vector(0,-1){0.5}}
\put(-0.15,-0.7){\makebox(0,0)[r]{no}}
\put(1.0,0.0){\vector(1,0){2.0}}
\put(2.0,0.05){\makebox(0,0)[b]{yes}}
}

\newsavebox{\arrparall}
\savebox{\arrparall}(0,0){%
\setlength{\unitlength}{0.5cm}
\thicklines
\put(-3,1){\line(1,0){8.0}}
\put(5,1){\line(-1,-1){2.0}}
\put(3,-1){\line(-1,0){8.0}}
\put(-5,-1){\line(1,1){2.0}}
\put(0,-1){\vector(0,-1){1.0}}
}
\newsavebox{\arrparalla}
\savebox{\arrparalla}(0,0){%
\setlength{\unitlength}{0.5cm}
\thicklines
\put(-3,1){\line(1,0){8.0}}
\put(5,1){\line(-1,-1){2.0}}
\put(3,-1){\line(-1,0){8.0}}
\put(-5,-1){\line(1,1){2.0}}
}

\begin{document}
\markboth{K.~Zhu, X.~H.~Mo, C.~Z.~Yuan}{Determining relative phase
in $\psp,~\jpsi$ decays into baryon and antibaryon}

%
\catchline{}{}{}{}{}
%

\title{Determination of the relative phase in $\psp$ and
$\jpsi$ decays into baryon and antibaryon}

\author{Kai~Zhu, Xiao-Hu~Mo, and Chang-Zheng~Yuan}
\address{Institute of High Energy Physics, CAS, Beijing 100049, China}

\maketitle

\begin{history}
\received{16 May 2015}
\revised{Day Month Year}
\end{history}

\begin{abstract}

With the recent measurements of $\psp$ and $\jpsi$ decay into
octet-baryon pairs, we study the relative phase between the strong
and the electromagnetic amplitudes, and find a large phase by
fitting the present data. The fits take into account the details
of experimental effects, including energy spread and initial state
radiation. We also predict some branching fractions of
$\psp$ decays and the continuum production rates at the $\jpsi$
mass based on the relative phase and absolute amplitudes obtained
from the fits.

\end{abstract}

\ccode{12.38.Aw, 13.25.Gv, 13.40.Gp, 14.40.Gx}
\maketitle

\section{Introduction}

Studying the relative phase between the electromagnetic (EM) and strong decay amplitudes,
in addition to the magnitudes of them, provides us a new viewpoint to explore the
quarkonium decay dynamics. Till now, no theory can give a satisfactory explanation of the
origin of or a constraint on this relative phase, a better knowledge of it may lead to a
better understanding of the quarkonium decay dynamics. Experimentally, the charmonium
($c\bar{c}$) states $J/\psi$ and $\psp$ are especially suitable for such a
study. First of all, they decay at the charm energy scale. Comparing with
lighter resonances, these decays provide more comparable amplitudes between strong and EM
processes. Comparing with heavier resonances, they provides larger fractions of the final
states with simple topologies, then benefits to experimental analysis. Furthermore,
$J/\psi$ and $\psp$ are vector resonances and can be produced directly via $e^+ e^-$
collision, and huge data samples were collected at many experiments. This means that very
small statistical uncertainties or/and good experimental precisions can
be achieved in measuring the decay branching fractions.

Studies have been carried out for many $\jpsi$ two-body mesonic
decay modes with various spin-parities:
$1^-0^-$~\cite{dm2exp,mk3exp}, $0^-0^-$~\cite{a00,lopez,a11}, and
$1^-1^-$~\cite{a11}, and baryon antibaryon pairs~\cite{ann}. These
analyses reveal that there exists a relative orthogonal phase
between the EM and strong decay amplitudes in $\jpsi$
decays~\cite{dm2exp,mk3exp,a00,lopez,a11,ann,suzuki}.

As to the $\psp$, author of Ref.~\cite{suzuki} argues that the
only large energy scale involved in the three-gluon decay of the
charmonia is the charm quark mass, then one expects that the
corresponding phases are not much different between $\jpsi$ and
$\psp$ decays. There is also a theoretical argument which favors
the $\pm90^\circ$ phase~\cite{gerard}. This large phase follows
from the orthogonality of the three-gluon and one-photon virtual
processes. Experimentally, some
analyses~\cite{wymppdk,wymphase,besklks1,cleoppdk1} based on
limited $\VP$ and $\PP$ data indicate that the large phase is
compatible with the data.

Recently more measurements with much improved precision for baryon
and antibaryon ($\BBb$) final states have been presented by
CLEO~\cite{cleobbdk1},
BES~\cite{besbbdk1,besbbdka,besbbdkb,besbbdkc} and
BESIII~\cite{Ablikim:2012eu,Ablikim:2012bw} Collaborations, at on-
and off-resonance regions for $\psp$ and $\jpsi$. These results
provide a possibility for a more precise phase analysis.
Comparing with analysis of mesonic decays the formalism for
bayonic decay is a little more complicated, therefore in
Sec.~\ref{xct_fmlm} the formalism we adopted is depicted, special
attention is paid for experimental conditions, such as energy
spread and initial state radiation (ISR) which have
profound effect on the fit results. The fits and corresponding
results are presented in Sec.~\ref{xct_fsfit} followed by a
summary.

\section{Description of the formalism}\label{xct_fmlm}

A general prescription of the parametrization of the amplitude for
the mesonic decays is given in Refs.~\cite{Haber,Morisita:1990cg},
where the decay amplitudes are expressed in terms of the
$SU(3)$-symmetric and $SU(3)$-symmetry breaking coupling
strengths. The idea and technique of this scheme are extended to
describe the baryonic decays below. Then by virtue of
parametrization form, we obtain the Born cross sections for all
possible modes of octet-baryon pairs, as well as the observed
cross sections taking all the experimental details into account.

\subsection{Parametrization}

Under $SU(3)_{flavor}$ symmetry (the subscript ``$flavor$'' is
omitted hereafter for briefness) the baryons can be arranged in
singlet, octet, and decuplet irreducible representations:
 $$ {\mathbf 3} \otimes {\mathbf 3} \otimes {\mathbf 3}
 = {\mathbf 1}_A \oplus {\mathbf 8}_{M_1} \oplus {\mathbf 8}_{M_2}
 \oplus {\mathbf 10}_S~.$$
The subscripts indicate antisymmetric ($A$), mixed-symmetric
($M_1$, $M_2$) or symmetric ($S$) multiplets under interchange of
flavor labels of any two quarks. Each multiplet corresponds to a
unique baryon number, spin, parity, and its members are classified
by isospin, its third component, and strangeness. The ground octet
and decuplet states, denoted as $B_8$ and $B_{10}$, correspond to
$J^P = \frac{1}{2}^+$ and $\frac{3}{2}^+$, respectively.

In an $SU(3)$ symmetric world only the decays into final states
$B_8 \overline{B}_8$ and $B_{10} \overline{B}_{10}$ are allowed,
with the same decay amplitudes for a given decay family if the EM
contributions are neglected. Nevertheless, the $SU(3)$ symmetry
can be broken in several ways~\cite{a11}, so in the
phenomenological analysis both symmetry-conserved and
symmetry-breaking terms are to be included. In this article we
analyze the octet-baryon pair final states.

To describe the $SU(3)$ octet states, it is convenient to
introduce the matrix notations
 \beq
{\mathbf B}=
\left(\begin{array}{ccc}
\Sigma^0/\sqrt{2}+\Lambda/\sqrt{6} & \Sigma^+   & p    \\
\Sigma^-  & -\Sigma^0/\sqrt{2}+\Lambda/\sqrt{6} & n    \\
\Xi^-     & \Xi^0            & -2\Lambda/\sqrt{6}
\end{array}\right)
\label{dknbyn}
\eeq
 and
\beq
\overline{\mathbf B}=
\left(\begin{array}{ccc}
\overline{\Sigma}^0/\sqrt{2}+\overline{\Lambda/}\sqrt{6}
                    & \overline{\Sigma}^- & \overline{\Xi}^-    \\
\overline{\Sigma}^+ & -\overline{\Sigma}^0/\sqrt{2}+\overline{\Lambda/}\sqrt{6}
                                          & \overline{\Xi}^0    \\
\overline{p}        & \overline{n} & -2\overline{\Lambda}/\sqrt{6}
\end{array}\right)
\label{dknatbyn}
 \eeq
for octet baryons and antibaryons, respectively. Note the fact
that, to an extremely good approximation, the $\psp$ and $\jpsi$
are $SU(3)$ singlets. For the decay $\psp$ and $\jpsi \to \BBb$,
to derive the consequences of the $SU(3)$ symmetry, the $SU(3)$
multiplets containing $B$ and $\overline{B}$ are combined in an
$SU(3)$-invariant way. The only such a combination is
 \beq
{\cal L}^0_{eff} = g \mbox{ Tr} (\mathbf { B \overline{B}})~,
\label{leffsym}
 \eeq
where ``Tr'' represents the trace of the matrix and the effective
coupling constant $g$ is proportional to the decay amplitude.

Now turn to the $SU(3)$-breaking effects. Following the recipe
proposed in Ref.~\cite{Haber}, $SU(3)$-breaking effect is
simulated by constructing an $SU(3)$-invariant amplitude involving
three octets and by choosing one of the octets (called a
``spurion'' octet) to point in a fixed direction of $SU(3)$ space
particular to the desired breaking. Two types of $SU(3)$ breaking
are considered. First, the quark mass difference. The $SU(2)$
isospin symmetry is assumed, that is $m_u=m_d$; but $m_s \neq
m_u$,~$m_d$ and this mass difference between $s$ and $u$/$d$
quarks leads to an $SU(3)$-breaking effect. By writing the quark
mass term as
 $$ m_d(\overline{d}d+\overline{u}u)+m_s \overline{s}s =m_0
 \overline{q}q + \frac{1}{\sqrt{3}} (m_d-m_s) \overline{q} \lambda_8 q~,$$
where $q=(u,~d,~s)$; $m_0=(2m_d+m_s)/3$ is the average quark mass;
$\lambda_8$ is the 8th Gell-Mann matrix. It can be seen that this
$SU(3)$ breaking corresponds to a spurion pointing to the 8th
direction of the abstract space spanning by 8 Gell-Mann matrices.
Explicitly, the matrix ${\mathbf S}_{m}$ is introduced to describe
such a mass breaking effect
 \beq
 {\mathbf S}_{m}= \frac{g_m}{3} \left(\begin{array}{ccc}
1 &   &  \\
  & 1 &  \\
  &   & -2
\end{array}\right)~,
\label{smbrkmass}
 \eeq
where $g_m$ is effective coupling constant due to the mass
difference effect.

Second, the EM decay amplitude. The EM effect violates $SU(3)$
invariance since the photon coupling to quarks is proportional to
the electric charge:
 $$ \frac{2}{3} \overline{u} \gamma_{\mu} u -
 \frac{1}{3} \overline{d} \gamma_{\mu} d - \frac{1}{3} \overline{s}
 \gamma_{\mu} s = \frac{1}{2} \overline{q} \gamma_{\mu} \left[
 \lambda_3+\frac{\lambda_8}{\sqrt{3}} \right] q~.$$
The above expression indicates that the EM breaking can be
simulated by the spurion matrix ${\mathbf S}_{e}$ as follows
 \beq
{\mathbf S}_{e}= \frac{g_e}{3}
\left(\begin{array}{ccc}
2 &    &  \\
  & -1 &  \\
  &    & -1
\end{array}\right)~,
\label{smbrkcharge}
 \eeq
where $g_e$ is effective coupling constant due to the EM effect.

To build an $SU(3)$ invariant out of three matrices, there are two
different ways of combination~\cite{hGeorgi}:
 $$ \mbox{ Tr} (\mathbf { B \overline{B} S} )~,~~~\mbox{or}~~~
 \mbox{ Tr} (\mathbf {\overline{B} B S} )~. $$
These are conventionally combined further into combinations
involving the commutator and anticommutator of the two matrices
and called $F$- and $D$-type, respectively. Therefore, the most
general form of $SU(3)$ invariant effective Lagrangian for three
matrices is
 \beq
{\cal L}_{eff} = D \mbox{ Tr} (\{\mathbf{ B,\overline{B}}\} \mathbf{S} )
+F \mbox{ Tr} ([\mathbf{B,\overline{B}}] \mathbf{S} ) ~.
\label{efflag}
 \eeq
Notice that $\mathbf{S}$ can be either ${\mathbf S}_{m}$ or ${\mathbf
S}_{e}$, together with ${\cal L}^0_{eff}$ for symmetric part, the
synthetic Lagrange reads
 \beq
\begin{array}{rcl}
{\cal L}_{eff} &=& g \mbox{ Tr} (\mathbf{B \overline{B}})
+ d \mbox{ Tr} (\{\mathbf{ B,\overline{B}}\} \mathbf{S}_{e})
+ f \mbox{ Tr} ([\mathbf{B,\overline{B}}] \mathbf{S}_{e}) \\
 & & + d^{\prime} \mbox{ Tr} (\{\mathbf{ B,\overline{B}}\} \mathbf{S}_{m})
+ f^{\prime} \mbox{ Tr} ([\mathbf{B,\overline{B}}] \mathbf{S}_{m})~.
\end{array}
\label{totefflag}
 \eeq

With the expressions in Eqs.~\eref{dknbyn}, \eref{dknatbyn},
\eref{smbrkmass}, and \eref{smbrkcharge} for $\mathbf {B}$,
$\mathbf{\overline{B}}$, ${\mathbf S}_{m}$, and ${\mathbf S}_{e}$,
respectively, the parametrization forms for octet-baryon-pair
final state are worked out and presented in
Table~\ref{octetbynform}, where the coupling constants are recast
as $A=g$, $D=d g_e/3$, $F=f g_e$, $\Dp=d g_m/3$, and $\Fp=f g_m$,
following the conventions in Refs.~\cite{a11} and
\cite{Kowalski:1976mc}.

\begin{table}[phtb]
\tbl{Amplitude parametrization forms for decays of the $\psp$ or
$\jpsi$ into a pair of octet baryons (phase space is not
included). General expressions in terms of
$SU(3)$-symmetry-conserved ($A$), as well as symmetric and
antisymmetric charge-breaking ($D$, $F$) and mass-breaking terms
($\Dp$, $\Fp$).}
 {\begin{tabular}{ll} \toprule
  Final state    & Amplitude parametrization form  \\ \colrule
  $\ppb$         & $A+D+F-\Dp+\Fp$    \\
  $\nnb$         & $A-2D-\Dp+\Fp$     \\
  $\SSbp$        & $A+D+F+2\Dp$       \\
  $\SSbz$        & $A+D+2\Dp$         \\
  $\SSbm$        & $A+D-F+2\Dp$       \\
  $\XXbz$        & $A-2D-\Dp-\Fp$     \\
  $\XXbm$        & $A+D-F-\Dp-\Fp$    \\
  $\LLb$         & $A-D-2\Dp$         \\
  $\SzLb+\SbzL$  & $\sqrt{3}D$        \\
\botrule
\end{tabular}\label{octetbynform}}
\end{table}

Here is a remark concerning the treatment of charge conjugate
final states. Applying the operator for charge conjugation to a
baryon-antibaryon system,
 \beq
C|{B}_n \overline{B}_m \rangle =| \overline{B}_n {B}_m \rangle \left\{
\begin{array}{ll}
   = |{B}_n \overline{B}_m \rangle & \mbox{  for } n=m \\
\neq |{B}_n \overline{B}_m \rangle & \mbox{  for } n \neq m
\end{array} \right. ~,
\label{eq_cpbbnm}
 \eeq
generally leads to a different state. Charge
conjugate states will be produced with the same branching
fractions, therefore we adopt the convention that charge conjugate
states are implicitly included in the measurement of branching
fractions, and the parametrization in Table~\ref{octetbynform} has
followed such an convention.

\subsection{Born cross section}

For $\EE$ colliding experiments, there is the inevitable continuum
amplitude
 $$ 
 \EE \rightarrow \gamma^* \rightarrow hadrons
 $$ 
which may produce the same final state as the resonance decays do.
The total Born cross section therefore
reads~\cite{rudaz,wymcgam,Wang:2005sk}
 \beq
 \sigma_{B}(s)=\frac{4\pi \alpha^2}{3s}
   |\ag(s)+\aga(s)+\ac(s)|^2~{\cal P}(s)~,
\label{bornxc}
 \eeq
which consists of three kinds of amplitudes correspond to (a) the
strong interaction ($\ag(s)$) presumably through three-gluon
annihilation, (b) the electromagnetic interaction ($\aga(s)$)
through the annihilation of $c\overline{c}$ pair into a virtual
photon, and (c) the electromagnetic interaction ($\ac(s)$) due to
one-photon continuum process. Notice $\aga(s)$ corresponds to the
contributions from resonance, $J/\psi$ or $\psi'$ here, then it
will be much larger than $\ac(s)$ even both of them are via single
virtual photon process. The phase space factor ${\cal P}(s)$ is
expressed as
 \beq
 {\cal P}(s) = v (3-v^2)/2~, ~~v\equiv
 \sqrt{1-\frac{(m_{B1}+m_{\bar{B}2})^2}{s}}~,
 \eeq
where $m_{B1}$ and $m_{\bar{B}2}$ are the masses of the baryon and
anti-baryon in the final states, and $v$ velocity of baryon in the
center-of-mass system (CMS).

For the octet-baryon-pair decay, the amplitudes have the forms:
 \beq
\ac(s)=\frac{Y}{s}~,
\label{ampac}
 \eeq
 \beq
\aga(s)=\frac{3Y\Gamma_{ee}/(\alpha\sqrt{s})}
{s-M^2+iM\Gamma_t}~,
\label{ampap}
 \eeq
 \beq
\ag(s)=\frac{3X\Gamma_{ee}/(\alpha\sqrt{s})}
{s-M^2+iM\Gamma_t}~,
\label{ampag}
 \eeq
where $\sqrt{s}$ is the center-of-mass energy, $\alpha$ is the QED
coupling constant; $M$ and $\Gamma_t$ are the mass and the total
width of the $\psp$ or $\jpsi$, respectively; $\Gamma_{ee}$ is the
partial width to $\EE$. $X$ and $Y$ are the functions of the
amplitude parameters $A,D,F,\Dp$, and $\Fp$ listed in
Table~\ref{octetbynform}, viz.
 \beq Y=Y(D,F)~, \label{defy} \eeq
 \beq X=X(A,\Dp,\Fp) e^{i\phi}~. \label{defx} \eeq
The special form of $X$ or $Y$ depends on the decay mode, as
examples, for $\ppb$ decay mode, $X=A-\Dp+\Fp$ and $Y=D+F$ while
for $\XXbm$ decay mode, $X=A-\Dp-\Fp$ and $Y=D-F$, according to
the parametrization forms in Table~\ref{octetbynform}. In
principle, the parameters listed in Table~\ref{octetbynform} could
be complex arguments, each with a magnitude together with a phase,
so there are totally ten parameters which are too many for nine
octet-baryon decay modes. To make the following analysis
practical, and referring to the analyses of measonic decays, it is
assumed that there is no relative phases among the
strong-originated amplitudes $A$, $\Dp$, $\Fp$, and no relative
phase between EM amplitudes $D$ and $F$; the sole phase (denoted
by $\phi$ in Eq.~\eref{defx}) is between the strong and the
electromagnet interactions, that is, between $X$ and $Y$ as
indicated in Eqs.~\eref{defx} and \eref{defy}, where $A$, $D$,
$F$, $\Dp$, and $\Fp$ are treated actually as real numbers.

\subsection{Observed cross section}

In $\EE$ collision, the Born cross section is modified by the ISR
in the way~\cite{rad.1}
 \begin{equation}
\sigma_{r.c.} (s)=\int \limits_{0}^{x_m} dx
F(x,s) \frac{\sigma_{B}(s(1-x))}{|1-\Pi (s(1-x))|^2},
\label{eq_isr}
 \end{equation}
where $x_m=1-s'/s$. $F(x,s)$ is the radiative function which has
been calculated to an accuracy of 0.1\%~\cite{rad.1,rad.2,rad.3},
and $\Pi(s(1-x))$ is the vacuum polarization factor. In the upper
limit of the integration, $\sqrt{s'}$ is the experimentally
required minimum invariant mass of the final state particles. In
the following analysis, $x_m=0.2$ is used which corresponds to an
invariant mass requirement of greater than $3.3$~GeV ($2.8$~GeV)
for the $\psp$~($\jpsi$) analysis.

The $\EE$ collider has a finite energy resolution which is much
wider than the intrinsic width of narrow resonances such as the
$\psp$ and $\jpsi$. Such an energy resolution is usually a
Gaussian distribution:
 $$
G(W,W^{\prime})=\frac{1}{\sqrt{2 \pi} \Delta}
             e^{ -\frac{(W-W^{\prime})^2}{2 {\Delta}^2} },
 $$
where $W=\sqrt{s}$ and $\Delta$, a function of the energy, is the
standard deviation of the Gaussian distribution. The
experimentally observed cross section is the radiative corrected
cross section folded with the energy resolution function
 \begin{equation}
\sigma_{obs} (W)=\int \limits_{0}^{\infty}
        dW^{\prime} \sigma_{r.c.} (W^{\prime}) G(W^{\prime},W)~.
\label{eq_engsprd}
 \end{equation}

\begin{table}[bthp]
\tbl{Breakdown of experiment conditions correspond to different
detectors and accelerators. The data taking position is the energy
which yield the maximum inclusive hadronic cross section. The data
with star ($\ast$) are the equivalent luminosity calculated with
relation ${\cal L}=N_{tot}/\sigma_{max}$. }
 {\begin{tabular}{llcccc} \toprule
         &         &Center-of-mass  &Data Taking    & Total & Integrated    \\
Detector & Accelerator &  Energy Spread    &Position
                                                  & events & luminosity  \\
         &             & (MeV)      & (GeV)         &($\times 10^6$)
                                                                     & (pb$^{-1}$) \\ \colrule
BESIII  & BEPCII      & 1.112       & 3.097         & 225.3  & 79.6   \\
CLEO~\cite{YELLOWBOOK}
         & CESR        &   1.5      & 3.68625       & 3.08
                                                                     & 2.74        \\
         &             &   2.3     & 3.68633        & 3.08
                                                                     & 2.89        \\
         &             &   2.28     & 3.671           & $-$     & 20.7        \\
BESII~\cite{besscan1}
         & BEPC        &   1.3     & 3.68623       & 14.0 & 19.72        \\
         &             &   1.27     & 3.650           & $- $  & 6.42        \\
         &             &   0.85    & 3.09700       & 57.7& 15.89$\ast$    \\
MARKII  & SPEAR       &   2.40    & 3.09711       & 1.32  & 0.924$\ast$    \\
 DMII   & DCI         &   1.98    & 3.09711       & 8.6 & 5.053$\ast$    \\
FENICE   & ADONE       &   1.24    & 3.09704       & 0.15  & 0.059$\ast$    \\ \botrule
\end{tabular} \label{tab_expcdn}}
\end{table}

As pointed out in Ref.~\cite{wymcgam}, the radiative correction
and the energy spread of the collider are two important factors,
both of which reduce the height of the resonance and shift the
position of the maximum cross section. Although the ISR are the
same for all $\EE$ experiments, the energy spread is quite
different for different accelerators, even different for the same
accelerator at different running periods. As an example, for the
CLEO data used in this paper, the energy spread varies due to
different accelerator lattices~\cite{YELLOWBOOK}: one (for
CLEO~III detector) with a single wiggler magnet and a
center-of-mass energy spread $\Delta$=1.5~MeV, the other (for
CLEO-c detector) with the first half of its full complement (12)
of wiggler magnets and $\Delta$=2.3~MeV~\cite{cleoppdk1}. The two
$\Delta$'s lead to two maximum total cross sections 635~nb and
441~nb, respectively, which differ prominently from BESII value of
717~nb for $\Delta=1.3$~MeV~\cite{besscan1}. All these subtle
effects must be taken into account in data analysis. In the
following analysis all data were assumed to be taken at the energy
point which yields the maximum inclusive hadron cross sections
instead of the nominal resonance mass~\cite{wymcgam,wymhepnp}.
Some experimental details are summarized in
Table~\ref{tab_expcdn}, and they are crucial for the data fitting
preformed below.

\section{Fit to data}\label{xct_fsfit}

Since our analyses involve the experimental details as indicated
in the preceding section, some measurements are not adopted in the
following study due to the lack of necessary information of the
detectors and/or accelerators. In addition, the status of the
accelerators are also different, so the fits to $\psp$ and $\jpsi$
decays are discussed separately for the sake of clarity.

\subsection{$\psp$ decays}

The experiment measurements were reported in
Refs.~\cite{cleobbdk1,besbbdk1}
and~\cite{Bai:2000ye,Feldman:1977nj,Brandelik:1979hy}. The results
of Refs.~\cite{Feldman:1977nj} and~\cite{Brandelik:1979hy} were
presented four decades ago and only one branching fraction (for
$\ppb$) and two upper limits (for $\LLb$ and $\XXbm$) were given.
The results of Ref.~\cite{Bai:2000ye} were obtained based on the
data taken 15 years ago. Therefore, only the results acquired
recently are utilized, which are quoted in Table~\ref{tab_pspdt}.

\begin{table}[bthp]
\tbl{Data of $\psp \to \BBb$ decays from CLEO~\cite{cleobbdk1} and
BESII~\cite{besbbdk1}. The continuum data are all from CLEO and
have been scaled by a factor $f_s$ as discussed in the text. The
errors are statistical only.}
{\begin{tabular}{ccccc} \toprule
  Mode  & $N^{obs}$      & $N^{obs}$    & efficiency  & Detector   \\
        &  (peak)        & (continuum)  & (\%)      &\\ \colrule
$\ppb$  &$556.5\pm 23.3$   &$15.9\pm 4.0$ & 66.6   & CLEO\\
        &$1618.2\pm 43.4$  &              & 34.4   & BESII\\
$\SSbp$ &$~34.2\pm 5.86$   &$~~~0\pm 2.3$ & ~4.1   & CLEO\\
$\SSbz$ &$~58.5\pm ~7.7$   &$~~~0\pm 2.3$ & ~7.2   & CLEO\\
        &$~59.1\pm ~9.1$   &              & ~4.4   & BESII\\
$\XXbz$ &$~19.0\pm ~4.4$   &$~2.0\pm 2.7$ & ~2.4   & CLEO\\
$\XXbm$ &$~63.0\pm ~8.0$   &$~1.8\pm 2.7$ & ~8.6   & CLEO\\
        &$~67.4\pm ~8.9$   &              & ~3.9   & BESII\\
$\LLb$  &$203.5\pm 14.3$   &$~3.4\pm 2.9$ & 20.1   & CLEO\\
        &$337.2\pm 19.9$   &              & 17.4   & BESII\\
        \botrule
\end{tabular} \label{tab_pspdt}}
\end{table}

It should be noted that for the results from CLEO Collaboration, the number of continuum
($N_{con}$) is not subtracted from the signal events at the $\psp$ peak. The
continuum data are all from CLEO and scaled by a factor $f_s=0.2547$ 
for all decay modes. $f_s=0.2547$ is calculated taking into account the
differences in luminosity and efficiency, and $1/s^5$ correction~\cite{cleobbdk1}. The
scaled results are shown in Table~\ref{tab_pspdt}. In addition, the CLEO data were taken
at two distinctive running states of the accelerator, which corresponds to different
energy spread, so the data are treated separately. If denoting the number of events taken
at CLEOIII as $N_1$ and at CLEO-c as $N_2$, then \beq
\begin{array}{rcl}
N_1 &=& {\cal L}_1 \cdot \sigma^1_{obs} \cdot \epsilon~,\\
N_2 &=& {\cal L}_2 \cdot \sigma^2_{obs} \cdot \epsilon~.
\end{array}
\label{eq_twonm}
 \eeq
Here the efficiencies ($\epsilon$) at the continuum and resonance
are considered to be the same~\cite{cleobbdk1}, $\cal L$ is the
integrated luminosity of the data sample, and $N$ the number of
observed signal events. So one gets
 \beq
 N = ({\cal L}_1 \cdot \sigma^1_{obs} +
      {\cal L}_2 \cdot \sigma^2_{obs}) \cdot \epsilon~,
 \label{eq_nmtwo}
 \eeq
where $N=N_1+N_2$ is the total number of signal events in the two data
sets.

Chi-square method is used to fit the experimental data. The
estimator is defined as
 \beq
 \chi^2= \sum\limits_i \frac{[N_i -n_i(\vec{\eta})]^2}{(\delta N_i)^2}~,
 \label{chisqbb}
 \eeq where $N$ denotes the experimentally measured number of events while $n$ the
theoretically calculated number of events.  The sum runs over all the final
states at the $\psp$ peak and the measurements at the continuum energy. The the five
continuum channels other than $p \bar{p}$ are combined to increase the
statistics. The observed cross section is calculated according to Eq.~\eref{eq_engsprd},
which contains the parameters to be fit, such as $A$, $D$, $F$, $\Dp$, $\Fp$, and the
phase $\phi$. All these parameters are denoted by the parameter vector $\vec{\eta}$ in
Eq.~\eref{chisqbb}. It should be noticed that $n$ consists of two parts for CLEO data and
should be calculated by Eq.~\eref{eq_nmtwo}.

The scan for each parameter discloses the two
minima of $\phi$ with opposite sign, while all the other
parameters have the same values up to the significant digits
listed below:

 \beq
\begin{array}{rcl}
   \phi  &=&(-98 \pm 25)^\circ~,\mbox{ or } ~(+134 \pm 25)^\circ~~; \\
       A &=&~~2.857 \pm 0.066~~; \\
     \Dp &=& -0.055 \pm 0.044~~; \\
     \Fp &=&~~0.060 \pm 0.066~~; \\
      D  &=&~~0.142 \pm 0.033~~; \\
      F  &=&~~0.027 \pm 0.052~~.
\end{array}
\label{fitpsp}
\eeq

The phase determined from $\psp \to \BBb$ decay is fairly
consistent with the analysis for $\psp \to \kskl$~\cite{besklks1},
where $\phi$ is determine to be $(-82 \pm 29)^\circ$ or $(+121 \pm
27)^\circ$. Here the solution $-98^\circ$ is more favorable for
the universal assumption proposed in Ref.~\cite{Wang:2003zx}. The
results of Eq.~\eref{fitpsp} show that for $\psp \to \BBb$ decays
the $SU(3)$-symmetric amplitude ($A$) dominates while the other
amplitudes are weaker by at least one order of magnitude.

With the above fit results, the ratios of the branching fractions
$Br(\psp \to n \bar{n}) / Br(\psp \to p \bar{p})$ and $Br(\psp \to
\Sigma^0 \bar{\Lambda} + c.c.) / Br(\psp \to p \bar{p})$ are
predicted to be $1.31 \pm 0.14$ and $0.007 \pm 0.004$,
respectively. Till now, there is no signal reported in experiments
for $\psp \to n \bar{n}$ and $\psp \to \Sigma^0 \bar{\Lambda} +
c.c.$.  We propose to measure them at experiments such as BESIII.
Notice that even the branching fraction of $\psp \to \Sigma^0
\bar{\Lambda} + c.c$ is only about $4\times 10^{-6}$ as we
predicted, with an assumption of $8\%$ reconstruction efficiency,
about 80 events can be observed with the 450~M $\psp$ events
collected at BESIII.

\subsection{$\jpsi$ decays}

There are lots of measurements at $\jpsi$ region. However, many of
measurements were performed almost ten or twenty years
ago~\cite{Brandelik:1979hy}-\cite{Bai:1998fu}. The recent
experimental results were mainly from
BES~\cite{besbbdka,besbbdkb,besbbdkc} and
BESIII~\cite{Ablikim:2012eu,Ablikim:2012bw} Collaborations.
Besides the data from them, the data from MARKII~\cite{mrk2bbdk}
and DMII~\cite{dm2bbdka,dm2bbdkb} are adopted, since the numbers
of events from these two experiments are considerable large and
more information of distinctive decay modes are provided. All data
used in this analysis are summarized in Table~\ref{tab_jpsidt}.

\begin{table}[bthp]
\tbl{Data of $\jpsi \to \BBb$ decays. The errors are statistical
only.}
 {\begin{tabular}{cccc} \toprule
  Mode  & $N^{obs}$          & efficiency      & Detector     \\
        &  (peak)            & (\%)      &              \\ \colrule
$\ppb$  &$~63316 \pm  281$    & 48.53     & BESII~\cite{besbbdka} \\
        &$~~1420 \pm  ~46$    & 49.7~     & MARKII~\cite{mrk2bbdk} \\
        &$314651 \pm  561$    & 66.1~     & BESIII~\cite{Ablikim:2012eu} \\
$\nnb$  &$~35891 \pm  211$    & ~7.7~     & BESIII~\cite{Ablikim:2012eu} \\
$\LLb$  &$~~8887 \pm  132$    & ~7.59     & BESII~\cite{besbbdkb} \\
        &$~~~365 \pm  ~19$    & 17.6~     & MARKII~\cite{mrk2bbdk} \\
        &$~~1847 \pm  ~67$    & 15.6~     & DMII~\cite{dm2bbdka}   \\
$\SSbz$ &$~~1779 \pm  ~54$    & ~2.32     & BESII~\cite{besbbdkb} \\
        &$~~~~90 \pm  ~10$    & ~4.3~     & MARKII~\cite{mrk2bbdk} \\
        &$~~~884 \pm  ~34$    & ~9.70     & DMII~\cite{dm2bbdka}   \\
$\SSbp$ &$~399.9 \pm 26.7$    & 0.462     & BESII~\cite{besbbdkc} \\
$\XXbz$ &$~203.6 \pm 21.0$    & 0.280     & BESII~\cite{besbbdkc} \\
$\XXbm$ &$~~~194 \pm  ~14$    & 12.9~     & MARKII~\cite{mrk2bbdk} \\
        &$~~~132 \pm  ~12$    & ~2.20     & DMII~\cite{dm2bbdkb}   \\
$\SzLb+\SbzL$
        &$~~~542 \pm  ~32$    & ~4.68     & BESIII~\cite{Ablikim:2012bw}   \\
\botrule
\end{tabular} \label{tab_jpsidt} }
\end{table}

The minimization estimator for $\jpsi$ is similar to that of $\psp$ as defined in
Eq.~\eref{chisqbb}. However, for $\jpsi$ data there is only limited
information about each detector, especially the integrated luminosity.  Therefore, it is
difficult to deal with all data consistently and accurately. To alleviate the systematic
biases among the data from different experiments, three scale factors are introduced. They
are normalized with respect to the BESII experiment and are floated in the fit. It should
be noted that no continuum data are available around $\jpsi$ mass, so we have less
constraint on the EM amplitudes than in the $\psp$ case.
The fitted parameters are listed as follows:
 \beq
\begin{array}{rcl}
  \phi  &=&(-85.9 \pm 1.7)^\circ~,\mbox{ or } ~(+90.8 \pm 1.6)^\circ~~; \\
       A &=&~~1.760 \pm 0.012~~; \\
     \Dp &=& -0.067 \pm 0.006~~; \\
     \Fp &=&~~0.102 \pm 0.013~~; \\
      D  &=&~~0.181 \pm 0.005~~; \\
      F  &=&~~0.168 \pm 0.088~~; \\
  f_{mk2}&=&~~0.904 \pm 0.024~~; \\
  f_{dm2}&=&~~0.704 \pm 0.021~~; \\
  f_{bes3}&=&~~0.922 \pm 0.004~~.
\end{array}
\label{fitjpsia}
 \eeq
Here the three factors $f_{mk2}$, $f_{dm2}$, and $f_{bes3}$
reflect the possible systematic bias in MARKII, DMII and BESIII,
respectively, relative to the BESII experiment. The fit values
indicate that the inconsistencies of these experiments from that
of BESII vary from 10\% to 30\%. This effect is small on the
determination of the phase, and is ignored in the discussion
below.

The phase determined from $\jpsi \to \BBb$ decays is similar to that from $\psp \to \BBb$
in this analysis and the magnitudes of amplitudes are similar too. We also notice that our
results are consistent with those in Ref.~\cite{Ablikim:2012bw}, in which the ``reduced
branching ratio'' method~\cite{a11,LopezCastro:1994xw} was applied, and the
$\phi$ is determined to be $(+76 \pm 11)^\circ$. It should be emphasized that with the
``reduced branching ratio'' method the continuum contribution is simply subtracted from the
data on the resonance peak, the interference between them has not been considered properly
and it can only provide relative strengths of the different amplitudes.

With the EM amplitudes determined from the fit, one can calculate the
continuum production cross sections of all the final states listed in
Table~\ref{tab_jpsidt}. As a byproduct, we predict
\begin{equation}
\begin{array}{l}
 \sigma(\EE\to \ppb)=11.5 \pm 5.8 \ \mathrm{pb}, \ \ \ \ \ \sigma(\EE\to \nnb)= 12.3 \pm 1.5 \ \mathrm{pb}, \\
 \sigma(\EE\to \LLb)= 2.8 \pm 0.4 \ \mathrm{pb},  \ \ \ \ \ \ \sigma(\EE\to \SSbz)= 2.7 \pm 0.3
 \ \mathrm{pb}, \\
\sigma(\EE\to \XXbz)= 9.2 \pm 1.1 \ \mathrm{pb},  \ \ \ \   \sigma(\EE\to \SSbp)= 10.0 \pm
 5.1 \ \mathrm{pb},  \\
 \sigma(\EE\to \SzLb + \SbzL)= 8.3 \pm 1.0 \ \mathrm{pb} 
\end{array}
\end{equation}
at a center-of-mass energy corresponding to the $\jpsi$ mass; while the cross sections of
$\sigma(\EE\to \SSbm) $ and $\sigma(\EE\to \XXbm)$ are at a few fb level. These
can be tested with the data samples at the BESIII experiment.

\section{Summary}\label{xct_sum}

The relative phase between the strong and the EM amplitudes of the
charmonium decays is studied based on the recent experimental data
of $(\psp,~\jpsi)\to \BBb$ decays.

For $\psp$ decays the phase is found to be $(-98 \pm 25)^\circ$ or
$(+134 \pm 25)^\circ$ while for $\jpsi$ decays the phase is fitted
to be $(-85.9 \pm 1.7)^\circ$ or $(+90.8 \pm 1.6)^\circ$. The
relative phases are similar between $\psp$ and $\jpsi$ decays into
baryon and anti-baryon final states, also are consistent with
previous results with meson final
states~\cite{dm2exp,mk3exp,a00,lopez,a11,ann,suzuki,wymppdk,wymphase,besklks1,cleoppdk1},
that should be updated with recent CLEO-c and BESIII measurements.
For the phase study in this work, the detailed experimental
conditions, such as energy spread and ISR, are taken
into account. However, due to the limited precision of the data,
only the strength of the dominate $SU(3)$-symmetric amplitude is
determined reasonably well. In order to fix all parameters which
describe the octet-baryon-pair decays, more accurate experimental
measurements are needed.

With the fit results, we also predict and propose to measure more
$\psp$ decays modes as well as the continuum production of the
baryon pairs at the $\jpsi$ mass region. Additional experimental
information will be helpful to draw a final conclusion on the
relative phase.

\section*{Acknowledgment}

This work is supported in part by National Natural Science
Foundation of China (NSFC) under contracts Nos. 11375206,
10775142, 10825524, 11125525, 11235011, 11475187; the Ministry of
Science and Technology of China under Contract Nos. 2015CB856701,
2015CB856706, and the CAS Center for Excellence in Particle
Physics (CCEPP).


\begin{thebibliography}{99}
\bibitem{dm2exp}J.~Jousset {\em et al.}, [DMII Collab.],
                Phys. Rev. D {\bf 41},  1389 (1990).

\bibitem{mk3exp}D.~Coffman {\em et al.}, [Mark III Collab.],
                Phys. Rev. D {\bf 38},  2695 (1988).

\bibitem{a00}M.~Suzuki, Phys. Rev. D {\bf 60},  051501 (1999).

\bibitem{lopez}G.~L\'{o}pez, J.~L.~Lucio M. and J.~Pestieau, arXiv:hep-ph/9902300.

\bibitem{a11}L.~K\"{o}pke and N.~Wermes,
                 Phys. Rep. {\bf 174},  67 (1989).

\bibitem{ann}R.~Baldini {\em et al.}, Phys. Lett. B {\bf 444},  111 (1998).

\bibitem{suzuki}M.~Suzuki, Phys. Rev. D {\bf 63},  054021 (2001).

\bibitem{gerard}J.-M.~G\'{e}rard and J.~Weyers,
                Phys. Lett. B {\bf 462},  324 (1999).

\bibitem{wymppdk} C.~Z.~Yuan, P.~Wang, X.~H.~Mo,
\Journal\PLB{567}{73}{2003}.

\bibitem{wymphase} P.~Wang, C.~Z.~Yuan and X.~H.~Mo,
\Journal\PRD{69}{057502}{2004}.

\bibitem{besklks1} J.~Z.~Bai {\em et al.} [BES Collaboration], 
\Journal\PRL{91}{052001}{2004}.

\bibitem{cleoppdk1} S.~Dobbs {\em et al.} [CLEO Collaboration],
\Journal\PRD{74}{011105}{2006}.  

\bibitem{cleobbdk1} T.K.~Pedlar {\em et al.} [CLEO Collaboration],
\Journal\PRD{72}{051108}{2005}.  

\bibitem{besbbdk1} M.~Ablikim {\em et al.} [BES Collaboration], 
\Journal\PLB{648}{149}{2007}. 

\bibitem{besbbdka} J.~Z.~Bai {\em et al.} [BES Collaboration], 
\Journal\PLB{591}{42}{2004}.

\bibitem{besbbdkb} M.~Ablikim {\em et al.} [BES Collaboration],
\Journal\PLB{632}{181}{2006}.


\bibitem{besbbdkc}
  M.~Ablikim {\it et al.}  [BES Collaboration],
  Phys.\ Rev.\ D {\bf 78}, 092005 (2008).

\bibitem{Ablikim:2012eu}
  M.~Ablikim {\it et al.}  [BESIII Collaboration],
  Phys.\ Rev.\ D {\bf 86}, 032014 (2012).

\bibitem{Ablikim:2012bw}
  M.~Ablikim {\it et al.}  [BESIII Collaboration],
  Phys.\ Rev.\ D {\bf 86}, 032008 (2012).

\bibitem{Haber}H.~E.~Haber and J.~Perrier, \Journal\PRD{32}{2961}{1985}.

\bibitem{Morisita:1990cg}N.~Morisita, I.~Kitamura and T.~Teshima,
Phys.\ Rev.\  D {\bf 44}, 175 (1991).


\bibitem{hGeorgi}H.~Georgi, ``{\em Weak interactions and modern particle
theory}'' (The Benjamin/Cummings Publishing Company, 1984), p65.

\bibitem{Kowalski:1976mc}H.~Kowalski and T.~F.~Walsh,
Phys.\ Rev.\ D {\bf 14}, 852 (1976).

\bibitem{rudaz}S.~Rudaz, \Journal\PRD{14}{298}{1976}.

\bibitem{wymcgam}P.~Wang, C.~Z.~Yuan, X.~H.~Mo and D.~H.~Zhang,
\Journal\PLB{593}{89}{2004}.

\bibitem{Wang:2005sk}P.~Wang, X.~H.~Mo and C.~Z.~Yuan,
Int.\ J.\ Mod.\ Phys.\  A {\bf 21}, 5163 (2006). 


\bibitem{rad.1} E.~A.~Kuraev and V.~S.~Fadin, Yad. Fiz. {\bf 41},
        733 (1985) [Sov. J. Nucl. Phys. {\bf 41},  466 (1985)].

\bibitem{rad.2} G.~Altarelli and G.~Martinelli, CERN {\bf 86-02}, 47 (1986);
        O.~Nicrosini and L.~Trentadue, Phys. Lett. B {\bf 196}, 551 (1987).

\bibitem{rad.3}F.~A.~Berends, G.~Burgers and W.~L.~Neerven,
        Nucl. Phys. B {\bf 297}, 429 (1988); {\it ibid.} {\bf 304}, 921 (1988).

\bibitem{YELLOWBOOK} CLEO-c/CESR-c Taskforces \& CLEO-c Collaboration,
 Cornell University LEPP Report No. CLNS~01/1742 (2001) (unpublished).

\bibitem{besscan1}BES Collaboration, J.~Z.~Bai {\em et al.},
\Journal\PLB{550}{24}{2002}.


\bibitem{wymhepnp}P.~Wang, C.Z.~Yuan and X.H.~Mo,
\Journal\HEPNP{27}{463}{2003}.

\bibitem{Bai:2000ye}J.~Z.~Bai {\it et al.}  [BES Collaboration],
Phys.\ Rev.\  D {\bf 63}, 032002 (2001). 

\bibitem{Feldman:1977nj} G.~J.~Feldman and M.~L.~Perl,
Phys.\ Rept.\  {\bf 33}, 285 (1977).

\bibitem{Brandelik:1979hy}  R.~Brandelik {\it et al.} [DASP Collaboration],
Z. Phys. C {\bf 1}, 233 (1979).

\bibitem{Peruzzi:1977pb} I.~Peruzzi {\it et al.},
Phys.\ Rev.\  D {\bf 17}, 2901 (1978).

\bibitem{Besch:1981ka} H.~J.~Besch {\it et al.},
Z.\ Phys.\  C {\bf 8}, 1 (1981).

\bibitem{mrk2bbdk} M.W.~Eaton {\it et al.} [MARKII Collaboration], Phys. Rev. D {\bf 29}, 804 (1984)

\bibitem{dm2bbdka} D.~Pallin {\it et al.} [DMII Collaboration],
Nucl. Phys. B {\bf 292}, 653 (1987).

\bibitem{dm2bbdkb} P.~Henrard {\it et al.} [DMII Collaboration],
Nucl. Phys. B {\bf 292}, 670 (1987).

\bibitem{Antonelli:1992ha} A.~Antonelli {\it et al.},
Phys.\ Lett.\  B {\bf 301}, 317 (1993).


\bibitem{Bai:1998fu} J.~Z.~Bai {\it et al.} [BES Collaboration],
Phys.\ Lett.\  B {\bf 424}, 213 (1998)


\bibitem{Wang:2003zx}P.~Wang, C.~Z.~Yuan and X.~H.~Mo,
Phys. Lett.  B {\bf 574}, 41 (2003).

\bibitem{LopezCastro:1994xw}
  G.~Lopez Castro, J.~L.~Lucio M. and J.~Pestieau,
  AIP Conf.\ Proc.\  {\bf 342}, 441 (1995)



\end{thebibliography}
\end{document}